\documentstyle[twoside,epsf,mypp]{article}

\newcommand{\myshowfig}[1]{\centerline{#1}}
\newcommand{\figcomment}[1]{}

\newcommand{\comment}[1]{}

\def\be{\begin{equation}}
\def\bel#1{\begin{equation}\label{eq:#1}}
\def\ee{\end{equation}}
\def\bea{\begin{eqnarray}}
\def\beal#1{\begin{eqnarray}\label{eq:#1}}
\def\eea{\end{eqnarray}}
\def\e#1{\label{eq:#1}}
\def\eqref#1{\ref{eq:#1}}

\def\eqnref#1{equation~(\ref{eq:#1})}
\def\eqnpref#1{eq.~[\ref{eq:#1}]}

\def\bfn{{\bf n}}
\def\Lcal{{\cal L}}
\def\Itilde{{\tilde I}}

\def\gsim{\lower 2pt \hbox{$\, \buildrel {\scriptstyle >}\over {\scriptstyle
\sim}\,$}}
\def\lsim{\lower 2pt \hbox{$\, \buildrel {\scriptstyle <}\over {\scriptstyle
\sim}\,$}}

\def\rg{R_g}
\def\rms{R_{ms}}

\def\ka{K$\alpha$}
\def\deg{$^\circ$}
\def\gamkep{\gamma_{\rm kep}}
\def\betkep{\beta_{\rm kep}}
\def\Omkep{\Omega_{\rm kep}}
\def\ems{{\rm em}}
\def\vpem{\vec{p}_\ems} 
\def\prec{\vec{p}_{\rm rec}}
\def\vem{\vec{v}_\ems}
\def\vrec{\vec{v}_{\rm rec}}

\def\tetr{\vec{e}_{\hat{r}}}
\def\teth{\vec{e}_{\hat{\theta}}}
\def\tetp{\vec{e}_{\hat{\phi}}}

\begin{document}

\noindent
\parbox{3in}{\footnotesize Submitted to The 
{\it Astrophysical Journal}\\ (November, 1997).}

\vspace{10pt}

\markboth{Rybicki \& Bromley}{Line Formation in a Relativistic Disk}
\pagestyle{myheadings}
\thispagestyle{empty}

\title{Emission Line Formation in a Relativistic Accretion Disk}
\author{George B. Rybicki and Benjamin C. Bromley}
\affil{Harvard-Smithsonian Center for Astrophysics}
\authoraddr{MS-51, 60 Garden Street, Cambridge, MA 02138}

\begin{abstract}

The observed profile of spectral lines from a relativistic accretion
disk can constrain parameters such as the disk geometry and the
rotation of the central black hole. The formation of the spectral line
in a disk generally has been modeled with simple assumptions such as
local isotropy of emission. Here we consider line formation in the
presence of velocity gradients induced by the differential flow in the
disk. In this case the emission can have anisotropy in the form of an
azimuthal dependence relative to the local principle axes of shear.
Since the physical conditions in a disk are uncertain in detail, we
investigate this effect with simple parameterized models based on
Sobolev theory to highlight the overall character of the changes in
the line profile. We find that velocity gradients generally cause a
relative increase of flux in the red wing, hence the inner radius of
the disk would be underestimated if the effect were not taken into
consideration. If the inner radius is used as a signature of black
hole rotation, as when the disk is not emissive within the marginally
stable circular orbit, then the inferred rotation would be
overestimated in cases where the emissivity of the disk has fairly
shallow fall-off with radius.  If the disk were emissive even within
the marginally stable orbit, then the local azimuthal anisotropy of
emission produces features in the line profile which distinguish
rotating from nonrotating black holes.

\comment{
It has been suggested that the observed profile of spectral lines from
a relativistic accretion disk can be used to constrain parameters such
as the rotation of the central black hole.  Such calculations have
generally assumed a simple model of the formation of a spectral line,
such that the intensity of emission is independent of the azimuthal
coordinate relative to the local principal axes of shear.  It is
pointed out that spectral line formation can be affected by the
anisotropic nature of the local shear field in the disk, and this can
in principle produce substantial effects in the observed profiles.
Because the relevant properties of the disk are uncertain in detail,
these effects are investigated here by introducing simple
parameterized models which are designed to give the general character
of the expected changes in the profile.  The effect of these
uncertainties on the possible measurement of black hole rotation is
discussed. We find that modifications to the line profile in optically
thick cases might provide features which would distinguish an unstable
freefalling disk around a nonrotating black hole from a stable
Keplerian disk in an extreme Kerr system.
}

\end{abstract}

\keywords{accretion, accretion disks -- black hole physics -- 
galaxies: active -- line: profiles}

\section{Introduction}\label{intro}


The treatment of line formation in disks is dependent on a number of
physical uncertainties and can be complicated if the medium is
optically thick in the line.  One complication
is the presence of velocity gradients within the disk, which can
affect the optical depth along a ray.  A larger velocity gradient
implies a smaller total optical depth because the same line absorption
is spread out over a larger bandwidth.  These velocity gradients are
highly anisotropic: for example, the local gradient of line of sight
velocity vanishes in a thin disk for rays which propagate radially
outward from the disk or which are tangential to the Keplerian
velocity field. The gradient reaches its maximum absolute value for
rays at 45 degrees to the radial or tangential direction (see Rybicki
\& Hummer \cite{RybHum83}). This implies that the emission along the
radial and tangential directions are suppressed relative to that along
the 45 degree directions. This can radically change the observed
profile of the line, since it is the tangential rays that
may contribute most strongly to the extreme Doppler shifts of the
line.

For the case of nonrelativistic disks, these velocity gradient effects
have usually been treated through use of the {\em Sobolev
approximation}, which assumes that the velocity gradients are
``large,'' in the sense that the typical Doppler shift of the line due
to velocity gradients is large compared with the intrinsic line width.
Rybicki \& Hummer (\cite{RybHum83}) discussed the validity of the Sobolev
approximation for nonrelativistic thin accretion disks.  Here the
critical question is whether the vertical scale of the relevant line
formation region is large enough for a single oblique ray to probe
parts of the disk with substantially different velocities, i.e.,
different enough to shift the profile through its own width.  Rybicki
\& Hummer (\cite{RybHum83}) concluded that the Sobolev approximation could
be marginally applicable for disks where the scale of line formation
was comparable to the entire disk thickness, but might be inapplicable
when the relevant line formation region was confined to a much thinner
layer near the top of the disk.  In their treatment of nonrelativistic
disks in cataclysmic variables, Horne \& Marsh (\cite{HorMar86}) avoided
making the Sobolev approximation of large velocity gradients to find
the escape probabilities, but their results were still limited by
the use of the escape probability method.

The effect of velocity gradients does not seem to have been previously
considered for extreme relativistic disks around Schwarzschild or
Kerr black holes.  Instead, these treatments in essence assume that
the emission from each disk element is isotropic, both in local
emission angle relative to the local disk normal and in local
azimuthal angle.

It is the purpose of this paper to evaluate the potential effects of
velocity gradients in the disk on the observed line profiles from
relativistic disks.  Because the detailed inclusion of these effects
depend on knowledge of many as yet unknown physical properties of the
disks, the development here can at most be a qualitative indication of
the types of phenomena to be expected.  We therefore find it
instructive to model the azimuthal dependence of the emergent emission
simply with the leading of a Fourier series as well as with the
full prediction of Sobolev theory.  This will allow us
to make some useful statements about the range of effects on the
observed line profiles due to velocity gradients and how such effects
might be degenerate with the effects of black hole rotation.

\section{The Line Formation Problem}\label{lfp}

In this section we review and extend the analytic theory of line formation in
differentially rotating disks. In \S2.1 we show how the observed line
can be related to the intensity field emerging from each local element
of the disk, including general relativistic effects. In \S2.2 we
discuss the line formation problem in the disk, using the
Sobolev theory.

 This will be followed
by the discussion of several approximation schemes, including the
general escape probability method and the Sobolev (large velocity
gradient) escape probability method.  Finally, these approximations
are compared with some accurate numerical solutions.

\subsection{The Specific Luminosity}\label{af}

Here we review and extend a number of results on the calculation of
the line spectrum from accretion disks.  Our notation will follow that
of Rybicki (\cite{Ryb70}) and Rybicki \& Hummer (\cite[hereafter
RH]{RybHum78}).  The purpose is to find formulas for the observed
monochromatic flux $F_\nu$ from a distant source (here, the accretion disk).
We limit the discussion to the case where the source subtends a very
small solid angle as viewed by the observer.

It is convenient to introduce the concept of the {\em specific
luminosity} $\Lcal_\nu(\bfn)$ of the source, which is defined as the
emitted energy at frequency $\nu$ per unit time, per unit frequency,
and per unit solid angle in the direction $\bfn$ of the observer.  In
Euclidean space, the monochromatic flux $F_\nu$ seen by the observer
at large distance $D$ is related to the specific luminosity by
\bel{th1}
    F_\nu = D^{-2} \Lcal_\nu(\bfn),
\ee
We remark that if the source is at cosmological distances, 
the observed flux can still be related to the specific luminosity, but
now through the more general formula,
\bel{th1'}
        F_\nu = (1+z)D_L^{-2} \Lcal_{\nu(1+z)}(\bfn),
\ee
where $z$ is the redshift and $D_L$ is the luminosity distance of the
source.

The calculation of $\Lcal_\nu$ can be facilitated by
constructing an observer's {\em plane of the sky} at the source location,
which is normal to the direction $\bfn$ of the observed radiation.
The specific luminosity can be expressed as an integral of the
specific intensity $I_\mu(\bfn)$ in direction $\bfn$ 
over this plane of the sky,
\bel{th2}
     \Lcal_\nu(\bfn) = \int I_\nu(\bfn)\,dA \ .
\ee
The plane of the sky does not have to be right at the
object, but should be sufficiently near so that both
the object and the plane are essentially at the
same distance from the observer.

For the non-relativistic case, the specific intensity field appearing
in the surface integral \eqnref{th2} can be found by tracing back
along each ray in a straight line until it intersects the disk, and
by using the invariance of specific intensity along rays.  This
means that the intensity appearing in Eq.\ \eqnref{th2}
is just equal to the corresponding intensity field emergent from the disk.

Both equations (\eqref{th1}) and (\eqref{th2}) can be directly applied to
relativistic disks, i.e., those surrounding black holes or other
compact objects for which general relativistic effects are
significant, providing that the plane of the sky is chosen to be
sufficiently far outside the gravitational field of the source
(ideally at infinity, but in practice typically at hundreds of
Schwarzschild radii).  However, new complications arise in relating
the specific intensity field on the plane of the sky to the field
emerging from the disk.  Two effects must be taken into account:
first, the rays now travel on curved paths in the strong gravitational
field of the source, and second, the invariant quantity along rays is
$I_\nu/\nu^3$, not $I_\nu$ itself.  These two effects can be expressed
by introducing the specific intensity $I'_{\nu'}$ emergent from the
disk corresponding to a particular ray in direction $\bfn$ in the
plane of the sky.  We define the redshift factor $g$ at each point of
the plane of the sky to be the ratio of the frequency $\nu$ to the
Doppler/gravitational shifted frequency $\nu'$ of the photon back
along the same ray as viewed in the comoving frame of the disk, that
is,
\bel{th3}
          g={\nu \over \nu'} \ .
\ee
The transformation of specific intensity is then,
\bel{th4}
        I_{\nu} = (\nu/\nu')^3 I'_{\nu'}=g^3 I'_{\nu/g} \ .
\ee
Therefore the generalization of \eqnref{th2} to relativistic
disks is,
\bel{th5}
         \Lcal_\nu(\bfn) = \int g^3I'_{\nu/g}\,dA.
\ee
The emergent specific intensity $I'_{\nu'}$ must be found from the
physics of the accretion disk itself.  For a thin disk, the radiation
field depends on only a fairly small patch of the disk, for which it
has been typical to use a locally nonrelativistic model as an
approximation.  

Emergent line radiation in the comoving frame of the disk is 
centered at the lab frequency of the line, which we shall denote $\nu_0$.
The line width is typically much smaller than the magnitude of the
subsequent frequency shifts due to Doppler/gravitation, so that it is
a good approximation to write
\bel{th6}
         I'_{\nu'} = \Itilde \delta(\nu'-\nu_0) \ ,
\ee
where $\Itilde$ is the integrated emergent line intensity in the
comoving frame of the disk,
\bel{th7}
          \Itilde = \int I'_{\nu'}\,d\nu' \ .
\ee
From well known transformation properties of $\delta$-functions we
have
\bel{th8}
          I'_{\nu/g} = g \Itilde \delta(\nu - g\nu_0) \ .
\ee
Using this in \eqnref{th5}, we obtain
\bel{th9}
          \Lcal_\nu(\bfn) = \int g^4 \Itilde \delta(\nu-g\nu_0)\,dA \ .
\ee
An equivalent form of this equation was given by Chen, Halpern \&
Filippenko (\cite{CheHalFil89}).

Two alternative forms of \eqnref{th9} will now be given.
The first is derived by introducing
the level curves of the $g$-function in the plane of the sky.  Let
$l$ be the arc length variable along each such level curve.
The element of area can be written
\bel{th10}
           dA = dl\,dw \ ,
\ee
where $dw$ is the width of the region between level curves corresponding
to $g$ and $g+dg$.   It is clear that
\bel{th11}
           dw= |\nabla g|^{-1} dg, \qquad\qquad dA = |\nabla g|^{-1} dg\,dl \ .
\ee
The delta function in \eqnref{th9} may be expressed as
\bel{th12}
         \delta(\nu - g\nu_0) = \nu_0^{-1} \delta(g-\nu/\nu_0) \ .
\ee
The integral of this delta function over $g$ then leads to a
formula for $\Lcal_\nu(\bfn)$ as a line integral,
\bel{th13}
   \Lcal_\nu(\bfn) = \nu_0^{-1} g^4\int \Itilde\, |\nabla g|^{-1} dl
\ee
over the curve of constant $g$ corresponding to $g=\nu/\nu_0$.  
(This curve might in general have several disconnected branches.)

We now derive a second alternative form for \eqnref{th9} that is more
convenient for practical calculations.  Let us
introduce a discrete frequency grid and define averages of
$\Lcal_\nu(\bfn)$ over frequency widths $\Delta\nu$ by
\bel{th14}
     {\tilde \Lcal}_\nu(\bfn)= {1 \over \Delta\nu} 
         \int\limits_{\nu-{\Delta\nu\over 2}}
                    ^{\nu+{\Delta\nu\over 2}} \Lcal_{\nu'}\,d\nu'
 ={1 \over \Delta\nu} \int g^4 \Itilde B\left({\nu-g\nu' \over \Delta\nu}
         \right)\,dA
\ee
where $B(x)$ is the ``boxcar'' function
\bel{th15}
   B(x)=\cases{1, & if $|x| \le {1 \over 2}$,\cr
              0, &  otherwise. \cr}
\ee
Replacing the integral by a sum over area elements $\Delta A_{ij}$,
we have,
\bel{th16}
     {\tilde \Lcal}_\nu(\bfn)
= {1 \over \Delta\nu} \sum g_{ij}^4 \Itilde_{ij} 
       B\left({\nu-g_{ij}\nu' \over \Delta\nu} \right)\,\Delta A_{ij}
\ee
To a consistent order of approximation, the factor $g_{ij}^4$ can be taken
outside the summation and replaced by $(\nu/\nu_0)^4$.
The boxcar function acts as a constraint on the summation, which leads to
our second alternative expression for the specific luminosity,
\bel{th17}
     {\tilde \Lcal}_\nu(\bfn)
= {1 \over \Delta\nu}\left( {\nu \over \nu_0}\right)^4 
       \sum_{|\nu-g_{ij}\nu_0| \le {1 \over 2}\Delta\nu}
             \Itilde_{ij}\,\Delta A_{ij} \ .
\ee
This form is well suited to calculating ${\tilde \Lcal}_\nu(\bfn)$
by accumulating contributions in separate bins corresponding to the
discrete frequency grid of resolution $\Delta\nu$.

In terms of algorithms for estimation of line profiles, \eqnref{th17}
suggests the following: a pixel image of the source, as seen by distant
observer, can be constructed wherein the pixels correspond to the area
elements $\Delta A_{ij}$. A calculation of the emergent intensity and
frequency shift of the radiation detected at each pixel then enables
the sum in \eqnref{th17} to be performed trivially. In practice, of
course, one iterates over pixels in the observer's sky plane and
simply accumulates the specific luminosity in each frequency bin. The
 difficulty in this procedure is modeling an image of
the source, which may involve relativistic effects such as
gravitational lensing. However, ray-tracing codes ---
photon trajectory integrators --- handle curved spacetime with
facility if not always with efficiency. They are of use not only in
determining the mapping of an observed pixel onto the source but can
also help in the calculation of the frequency shift $g$, which depends
on the 4-momenta of the emitted photons and the 4-velocity of the
source element. We discuss these issues in the next section.  But
first we must consider a second possible complication, that of
evaluating the local emergent intensity. We do so in the context of
Sobolev theory.

\subsection{Emergent Intensity and the Sobolev Approximation}
\label{ssect:sob}

The problem has now been reduced to determining the emergent
integrated intensities $\Itilde$ from the disk in a frame comoving
with the disk material.  Unfortunately, no definitive calculation of
the emergent intensities is possible at this time, since it depends on
highly uncertain disk parameters and even basic disk physics.
Therefore one is forced to make a number of simplifying assumptions
and approximations. We shall base our discussion on the
Sobolev escape probability formalism.

Most of the work on relativistic disks has assumed that the emergent
integrated intensity is some simple function of radial coordinate $r$,
perhaps with some dependence on the emergent polar angle $\theta_\ems$
relative to the local disk normal (sometimes called the
``limb-darkening law'').  The usual argument for the emergent
intensity to depend only the polar angle $\theta_\ems$ and not on the
azimuthal direction, is that each local patch of the disk is
essentially plane-parallel, with complete azimuthal symmetry.
However, from other work on accretion disks (see, e.g., Rybicki \& Hummer
\cite{RybHum83}) it is known that line emission can be strongly
influenced by local velocity gradients due to shear in the disk, and
this can induce azimuthal dependence in the emergent intensities by
causing frequency shifts in the line profiles of the same order or
larger than the line widths.

We first review the calculation for the simple case of a nonrelativistic,
Keplerian disk which is optically thin in the continuum but with
arbitrary optical thickness in the line.  (More details can be found
in \cite{RybHum83}.)  The usual cylindrical coordinate system, $r$,
$\varphi$, $z$ will be used.  The differential rotation of the disk
will be, by assumption, constant on cylinders with angular velocity
$\omega(r) \propto r^{-3/2}$, and the only nonzero components of the
rate of strain tensor $e_{ij}$ are $e_{r\phi_\ems}=e_{\phi_\ems
r}=(1/2)r\omega'=(3/2)\omega$.  Here the local
azimuthal angle $\phi_\ems$ is measured with respect to the tangential
velocity direction.  The velocity gradient along the ray $Q$ is given
by
\be
       Q= {3 \over 4} \omega \sin^2\theta_\ems \sin 2\phi_\ems \ . \e{thsob2} 
\ee

The emergent integrated intensity $\Itilde$ can be expressed as the
integral along the ray of the integrated emissivity $\epsilon$ times
the escape probability $P_{\rm esc}$,
\be
     \Itilde = \int \epsilon(r,z)P_{\rm esc}(r,z,\theta_\ems,\phi_\ems)\,dl \ . \e{thsob3}
\ee
Assuming that the quantities inside the integral depend only
weakly on $r$, then this integral can be converted to one over
the vertical direction,
\be
 \Itilde =\sec\theta_\ems \int 
        \epsilon(r,z)P_{\rm esc}(r,z,\theta_\ems,\phi_\ems)\,dz \ .  \e{thsob4}
\ee
in which $r$ appears only parametrically.  

In the special case of an optically thin line, the escape probability
is unity and we have
\be
         \Itilde = \sec\theta_\ems \int \epsilon(r,z)\,dz \ . \e{thsob5}
\ee
Thus the optically thin case shows strong ``limb-brightening,'' (the
emergent radiation is more intense for grazing emergence than for
normal emergence), but there is no dependence on the azimuthal angle
$\phi_\ems$ for this case.

The opposite case is for a line that is optically thick at least up to
some height $|z|=L$ from the disk midplane.  For $|z| > L$ we 
assume that the region rapidly becomes optically thin in the line
and does not contribute significantly to the emission.  The escape
probability in optically thick cases can be complicated by global
transfer effects, and even the formula \eqnref{thsob3} may not be
immediately useful because the emissivity itself may depend on the
global transfer problem.  However, for the case of large velocity
gradients, it turns out that the transfer is quite local, and the
theory is considerably simplified.  The line optical thickness
is given by the Sobolev optical depth
\be
     \tau_{\rm S} = {\kappa \over v_{\rm th} |Q|} \ . \e{thsob6}
\ee
where $\kappa$ is the integrated line opacity, and $v_{\rm th}$ is the
thermal Doppler velocity of the emitting ion.
The escape probability for
optically thick line is to a good approximation given in terms of
the Sobolev optical depth by the simple formula,
\be
      P_{\rm esc} = {1 \over \tau_{\rm S}} \ .    \e{thsob7}
\ee
The emergent integrated intensity is then found by combining Eqs.\
\eqref{thsob4}, \eqref{thsob6}, and \eqref{thsob7},
\be
   \Itilde = |Q|\sec\theta_\ems\int_{-L}^{L} {S v^{-1}_{\rm th}}\,dz \ ,
 \e{thsob8}
\ee
where $S=\epsilon/\kappa$ is the line source function.  
To apply this theory to thin relativistic disks, one must interpret
$Q$ to be $c$ times the gradient of the redshift factor along the
ray.

For the non-relativistic Keplerian disk, $|Q|$ may be replaced by
\eqnref{thsob2}, which gives
\be
         \Itilde = {3 \over 4} \omega
          \sec\theta_\ems \sin^2\theta_\ems \, |\sin2 \phi_\ems |
         \int_{-L}^{L} {S v^{-1}_{\rm th}}\,dz  \ . \e{thsob9}
\ee
Thus the angular distribution of the emitted line radiation at a
particular patch of the disk is proportional to $\sec\theta_\ems
\sin^2\theta_\ems \,|\sin 2\phi_\ems |$.
Although this result has been derived under a set of rather special
assumptions, it clearly demonstrates the potentially strong dependence
of the emission on the azimuthal angle $\phi_\ems$, through the factor
$|\sin2\phi_\ems|$.  This factor vanishes for the radial and tangential
directions (for which the line-of-sight velocity gradient vanishes),
and is maximum for the $\phi_\ems=\pm 45$\deg lines.

Technically, the Sobolev approximation breaks down when the
line-of-sight velocity gradient becomes small, so it is not strictly
true that the emergent intensity actually vanishes for radial and
tangential rays, but it is clear that the intensity will be reduced
relative to the lines where $|\sin 2\phi_\ems|=1$.  For this reason, and
because of other physical uncertainties in the theory, we shall
parameterize the azimuthal dependence of the emergent integrated
intensity using a Fourier cosine series with the same symmetry as 
$|\sin 2\phi_\ems|$,
\bel{Iser}
\Itilde = \Itilde_0 \sec\theta_\ems 
          + \Itilde_1 \sec\theta_\ems\sin_\ems^2\theta 
  \sum_{n = 0}^\infty a_n \cos(4n\phi_\ems)
\ee
where the first term on the righthand side corresponds the emissivity
of the disk in the absence of velocity gradient effects.  To
illustrate the effect that velocity gradients can have on emission
line profiles, we will take consider an emergent intensity constructed
with a gradient-free form and a first-order ``correction'' for
azimuthal dependence given by the $n = 1$ term.  Note that we recover
\eqnref{thsob9} by setting $\Itilde_0 = 0$ and
\bel{sxsob}
a_n = \left\{ 
        \begin{array}{ll} 
          \frac{4n}{\pi(1 - 4n^2)} & n \neq 0\ ; \\
          0                & n = 0 \ . 
        \end{array}
      \right.
\ee

The Sobolev approximation breaks down when line-of-sight velocity
gradients are small, since the local thermal broadening allows some
line photons to escape even along paths of zero velocity gradient.
This occurs in the present situation for a small set of emission
angles near the zero-gradient directions.  Although our Sobolev
calculations are technically wrong at these angles, we expect that the
overall solution will not be too much affected.  Our first order
angular expansion method errs in the opposite sense by giving too much
emission along those directions, and indicates that the
zero-gradient problem does not affect our general conclusions.

In order to evaluate the effect of small velocity gradients properly,
other, more exact, treatments of line formation should be used.  Such
a treatment for a nonrelativistic accretion disk was presented by
Horne \& Marsh (1986), who found the total emergent intensity without
use of the Sobolev approximation.  They were able to carry out most of
the calculation analytically by making several simplifying assumptions
about the structure of the disk, such as isothermal vertical structure
and constant line source function.  The resulting model of emergent
intensity has parameters which include a degenerate combination of the
height and temperature of the disk: the height regulates the breadth
of line emission from the range of Doppler shifts along a given line
of sight, while the temperature governs the intrinsic breadth of the
line at any given point in the disk. The ratio of these line breadths
determines the extent to which velocity gradients can affect the
emergent intensity --- the limit of zero temperature is the domain of
the Sobolev approximation.

Making analogous assumptions about the vertical structure and the
temperature as a function of position in the disk, it would be
straightforward to evaluate the emergent flux with a modified Horne \&
Marsh model.  However, the purpose of the present work is to estimate
the importance of the effect of velocity gradients in relativistic
disks where the run of temperature is not well-constrained, and the
Sobolev theory and the first-order correction term in equation~(27)
seem better suited to this preliminary analysis.  We expect that
our parameterization will at least qualitatively reproduce the effects
of velocity gradients given by more detailed models, which we leave
for future investigations.

\section{Relativistic Considerations}

In this section we consider aspects of line formation which are
specific to relativistic thin accretion disks.  We consider cases
where disk material is in stable circular orbit (Keplerian in a
relativistic sense), or, if any material exists within the marginally
stable circular orbit at radius $\rms$, it is assumed to be in
freefall onto the black hole, with energy and angular moment
corresponding to the marginally stable orbit (see Cunningham
\cite{Cun75}; also Reynolds \& Begelman \cite{ReyBeg97}). All
material, whether in circular orbit or in freefall, is taken to lie in
a slab whose thickness is everywhere small compared to the distance to
the black hole.  If the black hole itself has non-zero angular
momentum then we assume that the disk in is the equatorial plane. This
last assumption is justified by the Bardeen-Petterson
(\cite{BarPet75}) effect which causes a viscous disk to stabilize in
the equatorial plane of the rotating hole.  With these specifications
for the relativistic accretion disk system we proceed to calculate
quantities relevant to line formation.

\subsection{Ray Tracing}

As described in \S\ref{af}, the calculation of a line profile may amount to 
generating an image of the source, and binning up the image's pixels
according to frequency, weighted by pixel area, emergent intensity,
and the fourth power of the frequency shift. We can impose some model
for the emissivity over the face of the disk and apply corrections
to the emergent intensity, $I$, if limb darkening or velocity gradients
are present. What remains in the calculation is the mapping of
the observer's pixels to the source, the frequency shift,
and local emission angles upon which $I$ may depend.
A ray tracing code can provide the map from pixel to source, and
its output, photon 4-momenta, can be used in conjunction with a
model for the 4-velocity field of the source to obtain both $g$
and local emission angles.

Here, we obtain pixel images of relativistic thin disks from the code
described by Bromley, Chen \& Miller (\cite{BroCheMil97}).  It is a
general-purpose geodesic solver which has an incarnation as a ray
tracer when given photon positions and momenta on input. The code
integrates photon trajectories back in time from a distant observers'
sky plane until the photon intersects the disk. This is standard
procedure for ray tracing since we always know where the desired
photon trajectories end up (at the pixel array) but we may have only
guesses as to where these trajectories might originate. Note that
mapping from the source to the observer's sky plane is reasonable even
in geometrically flat spaces.  The case of image rotation is an
example; if we map pixels in a source image to a (rotated) target
image we may end up with target pixels which never
receive a photon. This is never a problem if we map from image to
source.

An output image of our code includes a list of pixels, their
coordinates in the observer's sky plane, and the location and
4-momentum of the photons at the point of intersection with the disk.

As is standard in the literature, we work in Boyer-Lindquist
coordinates, $(t,r,\theta,\phi)$, wherein the metric about a Kerr
black hole of mass $M$ and specific angular momentum $a$ takes the
form
\bel{kerrmetric}
      ds^2 = -\chi^2 dt^2 + \Psi^2( d\phi^2 - \omega dt)^2 +
             \frac{\rho^2}{\Delta} dr^2 +\rho^2 d\theta^2
\ee
with
\bea
&
     \chi^2 =  \frac{\Delta\rho^2}{A} \ , \ \ \ 
     \Psi^2 = \frac{A\sin^2\theta}{\rho^2} \ , 
& \nonumber \\
&
     A = (r^2+a^2)^2-a^2\Delta\sin^2\theta \ , 
     \ \ \ 
     \Delta = r^2+a^2-2Mr \ ,
& \\
&
      \rho^2 = r^2+a^2\cos^2\theta \ ,
      \ \ \ {\rm and} \ \ \ 
      \omega = \frac{2 M r a}{A} \ .
& \nonumber 
\eea
The contravariant 4-momentum of a photon with unit energy then 
becomes (Kojima \cite{Koj91})
\bel{4mo}
     (p_t,p_r,p_\theta,p_\phi) = 
          (-1,\pm\sqrt{R}/\Delta,\pm\sqrt{\Theta},\lambda)
      \ ,
\ee
where
\be
     R = (r^2+a^2-\lambda)^2-\Delta[(\lambda-a)^2+\eta] \ ,
\ee
and
\be \Theta = \eta^2 + \cos^2\theta \left(a^2 -
     \frac{\lambda^2}{\sin^2\theta}\right) \ .  
\ee 
The constants of motion, $\eta$ and $\lambda$, along with location
$(r,\theta,\phi)$ of the photon at some point in time, fully specify
the trajectory.

The photon coordinate energy in a local material frame follows from
projecting the 4-momentum $\vec{p}$ onto the material 4-velocity
$\vec{v}$ (the time basis vector of the material frame).  The
relative frequency of a photon exchanged between arbitrary material
frames is thus
\bel{gdef}
g \equiv 
%
%
\frac{-\prec \cdot \vrec} {-\vpem \cdot \vem} \ , 
\ee
where the subscripts distinguish emitter and receiver of the photon.  For the
distant observer at rest with respect to the black hole we set the receiver 
4-velocity to
\be
\vrec = \partial_t \ .
\ee
The 4-velocity of the emitter depends on the
nature of mass flow in the disk. In a Keplerian disk, we find
\bel{4vkep}
\vec{u}_{em} = \frac{\gamkep}{\chi}
               \left( \partial_t + 
               \Omkep \partial_{\phi}
               \right) \ ,
\ee
where
\be
\Omkep = M^{1/2}/\left( r^{3/2} + M^{1/2}a \right) 
\ee
is the coordinate angular velocity of a circular orbit and
$\gamkep$ is the Lorentz factor of the orbit with
speed
\be
\betkep = \frac{\Psi}{\chi}
 (\Omkep - \omega)
\ee
as measured in the locally nonrotating frame defined by Bardeen, Press
\& Teukolsky (\cite{BarPreTeu72}; see also Novikov \& Thorne
\cite{NovTho73}).  The relative frequency of a
photon from the disk detected by a distant observer is
thus
\bel{grelkep}
g = \frac{1}{\chi\gamkep (1 - \Omkep\lambda)} \ .
\ee

The case of a non-Keplerian disk, where the
emitters are in freefall from the minimum stable circular orbit
has been considered by Cunningham (\cite{Cun75}; details may be found
in the appendix therein).

The local emission angles, expressed in spherical polar coordinates, come from
projecting $\vpem$ onto the spatial basis vectors in the local emitter
frame:
\bel{emangs}
\cos\theta_\ems = -\frac{\vpem \cdot \teth}{\vpem \cdot \vem} \ ,
\ \ \ 
\sin\theta_\ems\cos\phi_\ems = 
  -\frac{\vpem \cdot \tetr}{\vpem \cdot \vem} \ ,
\ \ \
\sin\theta_\ems\sin\phi_\ems =
  -\frac{\vpem \cdot \tetp}{\vpem \cdot \vem} \ .
\ee
Note that $\theta_\ems$ gives the local inclination angle of the
photon trajectory relative to the disk surface, while $\phi_\ems$ is the 
angle between the photon's projected path in the disk plane and the
line tangent to circular orbit. A value of $\phi_\ems = 0$ indicates
emission in the forward direction of the material orbit. 

In the Keplerian case, for example, the vectors in
equations~(\eqref{emangs}) are
\bea
&
    \vec{e}_r = \frac{\Delta^{1/2}}{r} \partial_r '
    \ \ \
    \vec{e}_\theta = \frac{1}{r} \partial_\theta \ , 
& \nonumber \\
&
    \vec{e}_\phi = \gamkep \left( \frac{\beta}{\chi} \partial_t +
                   \frac{\Psi + \betkep\chi\omega}{\chi\Psi} \partial_{\phi}
                  \right) \ .
&
\eea

\subsection{Local Velocity Gradients}

As discussed by RH and reviewed in \S\ref{ssect:sob}, the optical
depth in an emission line becomes large when the local emission angle
is aligned with a constant-velocity surface.  In the case of a
nonrelativistic, thin disk, these surfaces appear as cross-hairs, ``$+$'',
with one hair parallel to the direction of flow in the Keplerian
velocity field. When we proceed to the relativistic case, 
the constant velocity surface must be generalized to a surface
of constant frequency shift so as to take into account effects
of space-time curvature in addition to Doppler shifts.

We can map out the surfaces of constant frequency shift with a ray
tracing technique, simply tracking photons from an emitter and noting
the value of $g$ at a spray of neighboring points. This can be
performed with \eqnref{gdef} which gives the observed
frequency shift for any arbitrary emitter--receiver pair. In a
Keplerian disk, 
\bel{gmap}
g = \frac{\chi(r) \gamkep(r+\Delta r) [1 - \Omkep(r+\Delta r) \lambda]}
         {\chi(r+\Delta r)\gamkep(r) [1 - \Omkep(r) \lambda]} \ ,
\ee
where the emitter is located at $(r,\phi)$ and the receiver is at
radius $(r + \Delta r, \phi + \Delta\phi$). Note that the constant of
motion $\lambda$ is related to $\Delta\phi$.  The constant-frequency
surfaces then can be mapped out in the Boyer-Lindquist coordinates by
varying $\lambda$ and $\Delta r$. 

It is useful to consider the problem of mapping the observed frequency
shift evaluated at some small proper radius $\epsilon$ in the emitter
frame as a function of emission angle. For now we consider only
receiver locations which lie in the disk plane, but our conclusions will
apply for disks with finite thickness as long as the velocity field is
independent of altitude above the disk.  A prescription is to choose
$\lambda$, thereby fixing the photon 4-momentum (\eqnpref{4mo}). This
in turn sets the local emission angles, i.e., a unit vector along the
photon propagation path as expressed in terms of the orthonormal basis
of the emitter frame.  We can then step to the point a distance
$\epsilon$ away along this vector to evaluate the observed frequency
shift $g$.  To gauge the dependence of $g$ on local emission angle
$\phi_\ems$ in the small--$\epsilon$ limit, we may rewrite the photon
4-momentum and 4-velocity at the receiver in terms of a Taylor
expansion in $r$ about the radial coordinate of the source. Then,
\beal{gexp}
g & = & \frac{(\vec{p}+\Delta r \vec{p}^{(r)}) \cdot
                        (\vec{v}+\Delta r \vec{v}^{(r)})}
         {\vec{p} \cdot \vec{v}} \ ,
\nonumber\\
  & = & 1 +  
        \frac{\vec{p} \cdot \Delta r \vec{v}^{(r)}}
             {\vec{p} \cdot \vec{v}} +
        \frac{\Delta r \vec{p}^{(r)} \cdot \vec{v}}
             {\vec{p} \cdot \vec{v}} + 
        O(\Delta r^2)\ ,
\eea
where all 4-vectors are evaluated at $r$ and the superscript
denotes differentiation with respect to $r$.
Evidently the second term on the RHS of \eqnref{gexp}
is a Doppler term, since $\Delta r \vec{v}^{(r)}$ represents
a change in the velocity field, while the third term is
associated with gravitational redshifting of the photon.
In the case of a Keplerian disk, the third term vanishes
because the gravitational redshift gradient is orthogonal
to the velocity flow. Then, since $\Delta r \sim \sin\phi_\ems$
and $\vec{p} \cdot \vec{v}^{(r)} \sim \cos\phi_\ems$, we recover the
nonrelativistic result in the limit of small $\epsilon$:
\bel{gnonrel}
g - 1 \sim  \sin2\phi_\ems 
\ .
\ee
%

Deviations from this expression for finite $\epsilon$ cause
distortions in the sine-wave dependence of $g$ on local emission
angle, but the zero-crossings, and hence the constant-frequency-shift
surfaces, are preserved.

For the case of freefall orbits, similar results can be obtained
although in this case there are convergent flow lines and changes in
velocity along the direction of flow. Numerical calculations
nonetheless confirm that the cross-hair form for the zero-frequency-shift
surfaces is a reasonable approximation.


\section{Results: Line Profile Calculations}

We have reviewed all the ingredients necessary to estimate the effects
of velocity gradients on emission line profiles from relativistic
accretion disks. Thus armed we proceed to numerically calculate
profiles in the following three settings: 1) a stable Keplerian disk
with an inner radius at the marginally stable limit, $\rms = 6$~$\rg$
around a Schwarzschild black hole ($\rg = \sqrt{G M/c^2}$ is the
gravitational radius of the black hole); 2) a stable Keplerian disk
extending down to 1.25~$\rg$ around an extreme Kerr black hole; and 3)
a disk in freefall inside marginally stable orbit around a
Schwarzschild hole.  In all of these cases we take the intrinsic disk
emissivity $\Itilde_0$ (see \eqnpref{Iser}) to be a power law in radius with
index $\alpha$ in the range of -4 to 0. Throughout we seek the changes
that velocity gradients can cause in a line profile and we wish to
determine how these changes can affect the interpretation of the
profile if they are not taken into account. Specifically we are
interested to know if these effects cause confusion or highlight the
differences between profiles of the three systems listed above.

\begin{figure}[hp] 
\myshowfig{\epsfxsize=6.0in\epsfbox{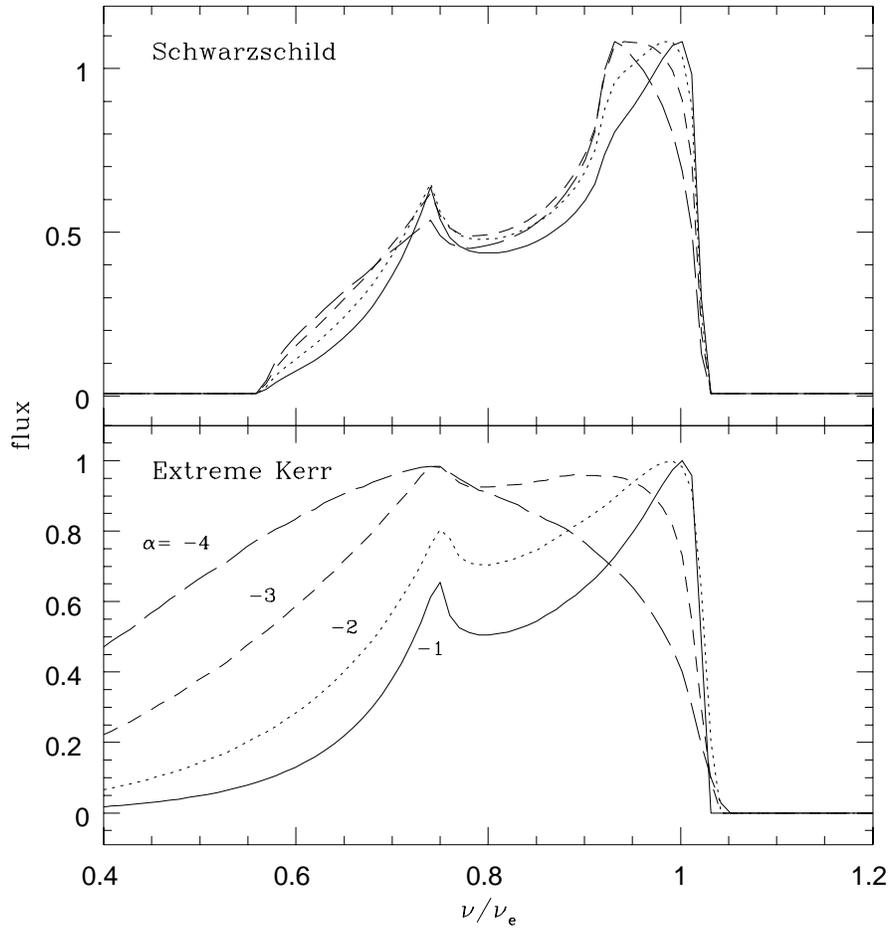}}
\caption{ 
Line profiles from disk models with locally isotropic, power-law
emissivity functions for various values of the index $\alpha$. The
upper plot corresponds to a Keplerian disk of radii between 6~$\rg$
and 12~$\rg$ around a nonrotating black hole. The lower plot
corresponds to a Keplerian disk about a Kerr black hole with
extremal angular momentum; the inner and outer radii of the disk
are 1.25~$\rg$ and 12~$\rg$. The inclination angle
in both cases is 30\deg.
}
\label{fig:lpex}
\end{figure}

We begin with an illustration (Fig.~\ref{fig:lpex}) of the dependence
of a line profile on the choice of power-law index $\alpha$. For the
moment we only consider Keplerian disks.  By definition, steeper power
laws enhance emission of the inner disk.  Note that in the
Schwarzschild system, with the Keplerian disk extending down to its
innermost stable orbit, the location of the blue peak must be above
$\nu/\nu_e = 0.72$ for any inclination angle $i$; this limit rises
with increasing $i$, so that at $i = 30$\deg, the blue peak is located
above 0.92, regardless of $\alpha$.  The trough between the two peaks
of the Schwarzschild case is a characteristic of this system. The
trough can be ``filled in'' to some degree by weighting outer disk
components more strongly, although the result is always to create a
stronger blue peak close to the rest frame frequency of the line.  In
contrast, the line profiles observed from a disk around an extreme
Kerr black hole can have a much shallower trough from emission of
sources below 6~$\rg$ (e.g., Laor 1991). This property provides a
signature of the Kerr system in addition to the extended red wing
(e.g., Laor 1991). However, the signature is strong only if the inner
disk component is heavily weighted, with $\alpha \lsim -2$.

Changes in the line profile from variation of other parameters, namely
the inclination angle and disk radii, have been considered in the
literature (Asaoka 1989\nocite{Asa98}; Laor 1991\nocite{Lao91}; Kojima
1991\nocite{Koj91}; Bromley, Chen \& Miller 1997\nocite{BroCheMil97}).
In qualitative terms, an increasing inclination angle broadens the
line profile and tends to strengthen the signatures of emission at
small disk radii.  The latter effect comes from the increased
importance of gravitational amplification at high inclination angle.
In a Schwarzschild system, when $\alpha \gsim -4$ and the radial
extent of the disk is finite, two Doppler peaks are generally evident
in the line profiles; the red peak directly reflects the outer disk
radius and the inclination angle while the precise location of the
blue peak depends on the index $\alpha$ as well as the inclination
angle and disk radii.  Kerr systems are similar except that distinct
Doppler peaks may no longer exist because the blue peaks from emitters
at small radii can be redder than the red peaks from sources at large
radii.

The simple power-law emissivity model just described is built on the
assumption that the sources in a disk are isotropic in their rest
frames, and that their distribution on a given disk annulus is
uniform.  The latter assumption is relaxed in studies of emission from
``hot spots'' on the disk (e.g., Bao 1992). The isotropy assumption
has also been relaxed for relativistic accretion disks, but only in
the context of limb darkening, that is, a dependence on the emission
angle $\theta_\ems$. In the present study, we consider the dependence
on azimuthal emission angle $\phi_\ems$ to determine how velocity
gradients might affect a disk line and if they can cause confusion in
the interpretation of a line as coming from a Schwarzschild or Kerr
system.  The details of radiative transfer are uncertain, inasmuch as
the physical structure of an accretion disk around a supermassive
black hole has yet to be determined. Thus, as promised, we use only
the highest amplitude ($n = 1$) term in the expansion for emergent
flux given in \eqnref{Iser}.  We hereafter refer to this term as the
``leading correction'' for velocity gradients.

\begin{figure}[hp] 
\myshowfig{\epsfxsize=6.0in\epsfbox{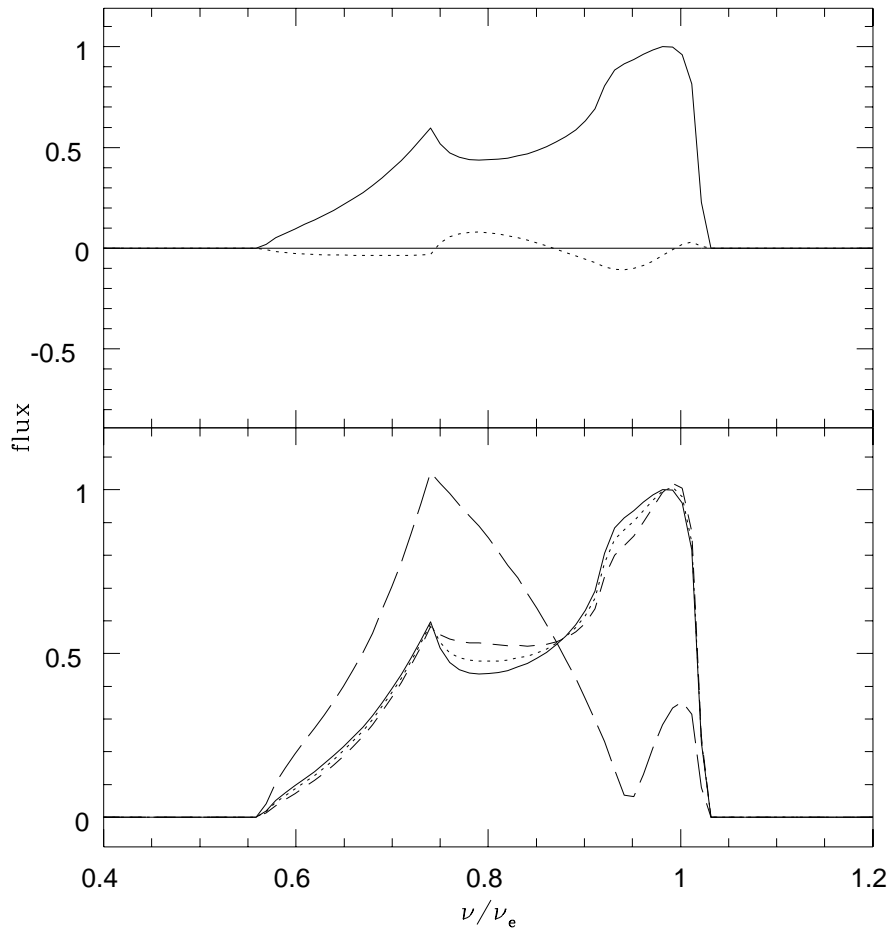}}
\caption{ Line profiles from a Schwarzschild disk model with a
power-law emissivity index $\alpha = -2$. Inclination angle and disk
radii are the same as in Figure~1. The upper plot shows the case of
isotropic emission (solid curve) and the flux contribution (dotted
line) from the first-order correction for the velocity gradient
effect. The lower plot shows the isotropic case and linear
combinations with the first-order contribution. The weights of the
first-order terms are 0.3 (dotted curve) and 0.5 (short dashed curve).
The prediction of the Sobolev theory is show as well (long dashed cyrve).
  }
\label{fig:lpsx}
\end{figure}

\begin{figure}[h] 
\myshowfig{\epsfxsize=6.0in\epsfbox{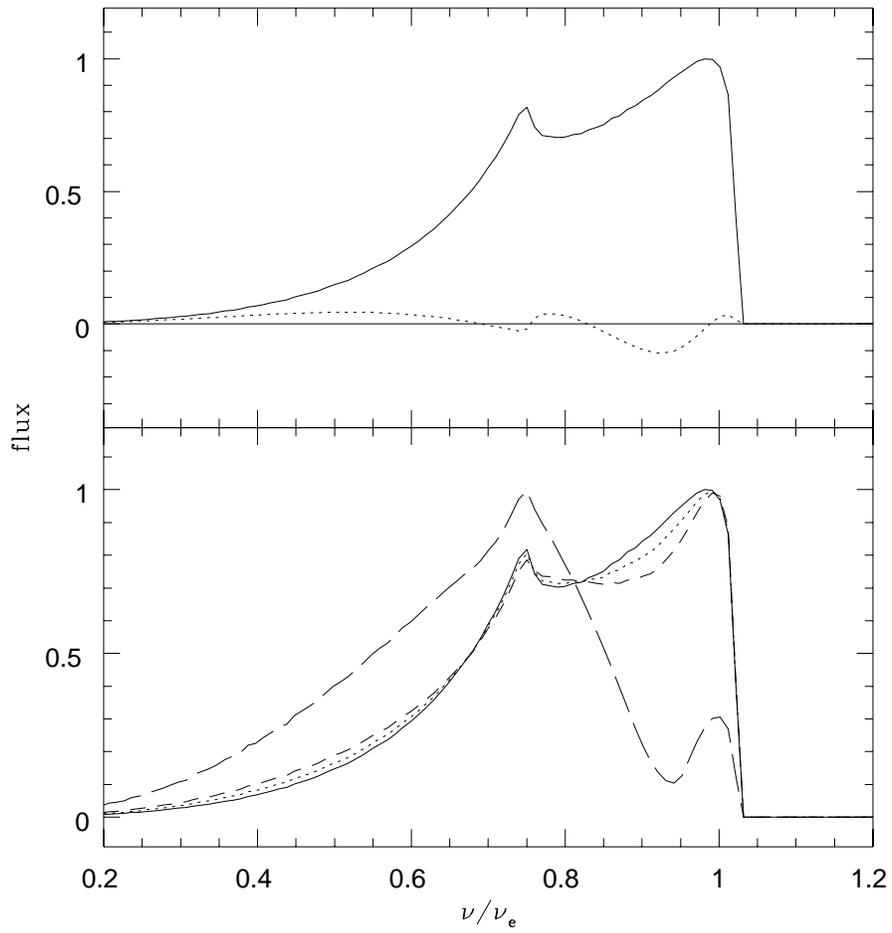}}
\caption{ Same as Figure 2 except for a Kerr disk with inner radius of
1.25~$\rg$. 
}
\label{fig:lpkx}
\end{figure}

Figure~\ref{fig:lpsx}
shows the effect of the leading correction to the line
profile from a Schwarzschild system.  Evidently, the trough in the
Schwarzschild profile is enhanced after the velocity-gradient effect
is included.  A similar enhancement is seen in the extreme Kerr system
(Fig.~\ref{fig:lpkx}). To neglect the velocity-gradient effect, if
it indeed is present in an observed emission line, is to overestimate
the depth of the central trough. Thus, confusion between 
Kerr and Schwarzschild systems may result.

\begin{figure}[h] 
\myshowfig{\epsfxsize=6.0in\epsfbox{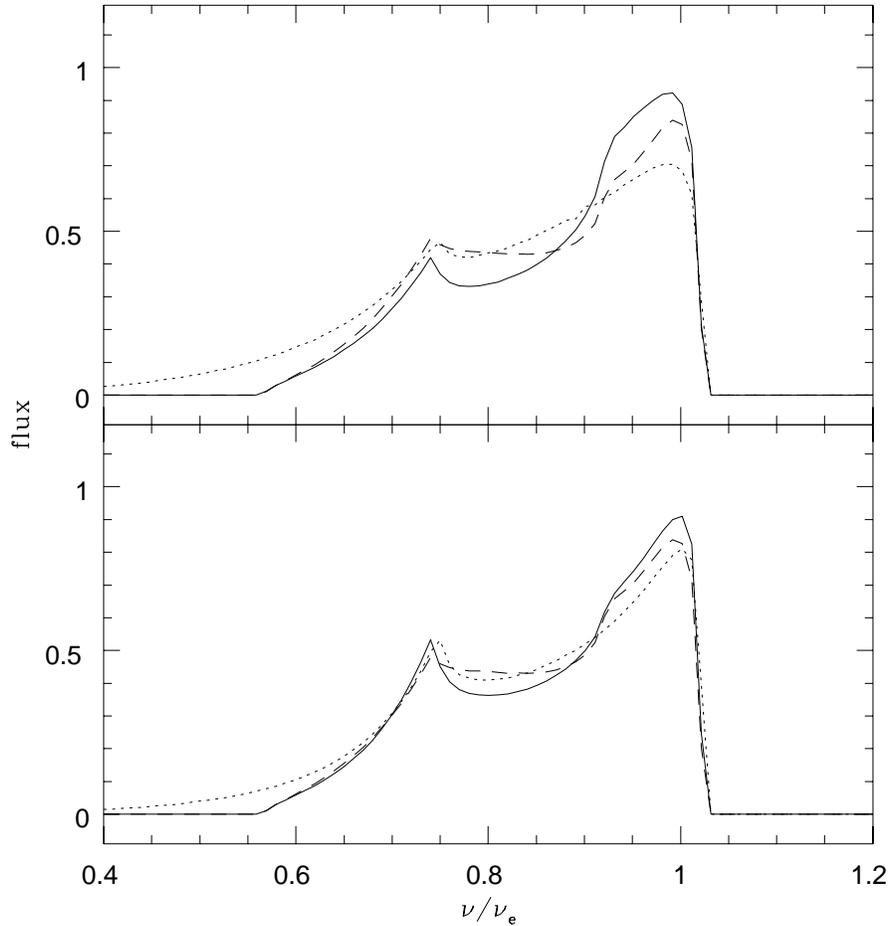}}
\caption{ Comparison of line profiles with and without first-order
velocity gradient corrections.  The upper plot shows profiles from
both Schwarzschild (solid curve) and Kerr (dotted curve) systems with
$\alpha = -2$ and the same disk parameters as in Figure~1. A profile
from the Schwarzschild system, corrected for the velocity gradient
effect to first order with limb-darkening factor of $\sin(\theta_e)$,
is also shown (dashed curve). In the lower plot, the same corrected
profile is shown, but now superimposed on profiles from Schwarzschild
and Kerr systems with $\alpha = -1$.  }
\label{fig:lpmx}
\end{figure}

This point is illustrated in Figure~\ref{fig:lpmx} wherein profiles
with and without velocity-gradient corrections are superimposed.  If
the data are noisy at the level of recent ASCA observations (e.g.,
Tanaka et al. 1995), then certain key features of a Kerr system such
as emission from small radii will not be directly detected. The form
of the line profile above $\nu/\nu_e \sim 0.6$, in particular the
depth of the trough between red and blue peaks, is thus crucial to the
identification of a Kerr system. In Figure~\ref{fig:lpmx} profiles
from a Schwarzschild system in which the velocity gradient is
important are compared with profiles from Kerr and Schwarzschild
systems in which the effect is irrelevant.  As illustrated, profiles
from a corrected Schwarzschild system may be misinterpreted as
evidence of an extreme Kerr model on the basis of a $\chi^2$
metric. However, the confusion can only occur if the emissivity index
is near -2 or greater, so that the inner Kerr disk is not weighted
heavily.

It is evident from Figures \ref{fig:lpex}--\ref{fig:lpmx} that if
one interprets a line profile as having a non-disk Gaussian component
(e.g., Iwasawa et al. 1996), then the excess blue peak caused by the
velocity gradient correction will no longer be associated with the
disk line and its removal will cause the profile to more closely
resemble a Kerr system.

\begin{figure}[h] 
\myshowfig{\epsfxsize=6.0in\epsfbox{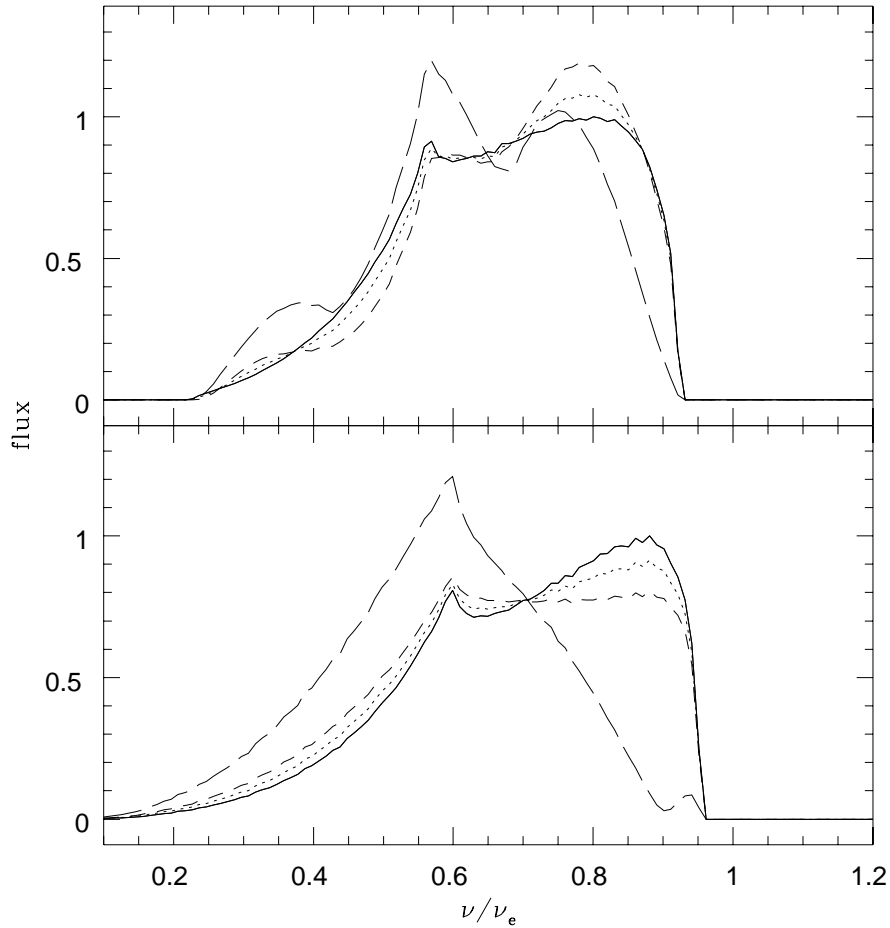}}
\caption{ Line profiles from disk material below 6~$\rg$. The upper
plot is for a disk in freefall from from the marginally stable orbit
down to a radius of 2.1~$\rg$ about a nonrotating black hole. The
lower plot corresponds to a disk around an extreme Kerr hole with an
inner radius of 1.25~$\rg$.  In both cases the inclination angle is
30\deg\ and the emissivity index is -2. The curves indicate different
levels of contribution from the velocity-gradient effect, as in 
Figure 2.
}
\label{fig:lpfx}
\end{figure}

In the case of a freefalling disk in a Schwarzschild system
significantly redshifted emission may be comparable to that seen in
the extreme Kerr case with a Keplerian disk.  All orbits are unstable,
as they are assumed to be infinitesimally perturbed inward from $\rms$
at 6~$\rg$.  Reynolds \& Begelman (\cite{ReyBeg97}) imply that the
line profiles from a freefalling disk and a Kerr system may be
indistinguishable and we confirm that result (see
Fig.~\ref{fig:lpfx}). Evidently the differences in orbital
trajectories make only negligible changes in the observed photon
redshifts, at least when integrated over the disk.  However, if the
local emission angles play a role, as in the velocity gradient effect,
then the trajectory differences become
important. Figure~\ref{fig:lpfx} illustrates that in some instances
when the inner disk, the velocity gradient effect can break the
approximate degeneracy in the profile shapes.

\section{Summary}

We have reviewed aspects of the line formation problem for a
relativistic accretion disk. Our new contribution is to include the
effects of velocity gradients on line luminosity when the disk has
some optical thickness in the line. From the illustrations given in
the previous section we can draw the following conclusions regarding
the use of line profiles to determine properties of the accretion
disk/black hole system from the shape of line profiles.  If disks are
stable, then the velocity-gradient effect can mimic an increase in the
emission from small radii.  When the inner disk is not too heavily
weighted, for example if emissivity index is -2 or more, then black
hole rotation might be spuriously inferred. If disks are emissive even
in freefall, then the effects of rotation are virtually invisible in
the line profiles. However, in this case the velocity-gradient effect
may help distinguish the Kerr and Schwarzschild systems, as it can
leave a different imprint on otherwise similar profiles.  This is
possible only if the inner disk is weighted heavily, as when the
emissivity index is less than -2.

The velocity gradient effect may have relevance to the iron \ka\
emission lines which have been recently studied (e.g., Tanaka et
al.~\cite{Tanetal95}) in data from Seyfert~1 galaxies. The \ka\ line
complex arises primarily from fluorescence. If an accretion disk is
cold, with material less ionized than Fe~XVIII, then it will be
optically thin in the line and no velocity gradient effect will be
observable. However, the effect might occur in the inner region of the
disk which may support higher ionization states that will resonantly
scatter emission line photons.  Thus the effect may be measurable,
offering a new probe of the local environment of the central
supermassive black hole.

\acknowledgements 

This work was supported by NSF grant PHY 95-07695.
The Cray supercomputer used in this research was provided through
funding from the NASA Offices of Space Sciences, Aeronautics, and
Mission to Planet Earth.

\end{document}